\documentclass[aps,pra,twocolumn]{revtex4}
\usepackage{graphicx}
\usepackage{hyperref}
\begin{document}

\title{Towards
experimental entanglement connection with atomic ensembles\\ in the
single excitation regime}

\author{Julien Laurat, Chin-wen Chou, Hui Deng, Kyung Soo Choi, Daniel Felinto, Hugues de Riedmatten, H. J. Kimble}

\address{Norman Bridge Laboratory of Physics 12-33, California Institute of
Technology, Pasadena, California 91125, USA}

\begin{abstract}
We present a protocol for performing entanglement connection between
pairs of atomic ensembles in the single excitation regime. Two pairs
are prepared in an asynchronous fashion and then connected via a
Bell measurement. The resulting state of the two remaining ensembles
is mapped to photonic modes and a reduced density matrix is then
reconstructed. Our observations confirm for the first time the
creation of coherence between atomic systems that never interacted,
a first step towards entanglement connection, a critical requirement
for quantum networking and long distance quantum communications.
\end{abstract}

\maketitle

\section{Introduction}

The distribution of entanglement between different parties enables
the realization of various quantum communication protocols, such as
quantum cryptography, dense coding and teleportation
\cite{chuang00,zoller05}. Such distribution can rely on entanglement
swapping, namely the teleportation of entanglement, which aims at
entangling two distant systems which never interacted in the past.
Important aspects of this striking feature have already been
demonstrated with independent sources of entangled light. In the
discrete variable regime, one can generate two independent pairs of
polarization entangled beams and subject a superposition of two of
the beams to a Bell-state measurement. The two remaining beams are
then projected into an entangled state \cite{pan98}. More recently,
unconditional entanglement swapping has been achieved for continuous
quantum variables of light \cite{jia04,takei05}.

However, to enable quantum communication over arbitrary long
distances, entanglement needs to be stored in matter systems. In the
quantum repeater architecture \cite{Briegel98}, entanglement is
distributed by swapping through a chain of spatially separated
entangled pairs of memories, leading to the possibility of scalable
long-distance communication. Connecting entangled matter systems is
thus a critical requirement for the practical realization of quantum
networks. Along this line, generation of entanglement between atomic
systems has been reported, including
entanglement of the discrete internal states of two trapped ions \cite%
{Turchette98}, long-lived entanglement of macroscopic quantum spins \cite%
{julsgaard01} and, more recently, heralded entanglement between atomic
ensembles in the single excitation regime \cite{chou05}. However, no
entanglement connection has been demonstrated so far with such matter
systems. In this paper, we present our work towards entanglement connection
of atomic ensembles and demonstrate for the first time the transfer of
coherence between two atomic ensembles which never interacted.

The paper is organized as follows. Section \ref{2} gives a brief
overview of our matter building block, namely an atomic ensemble in
the regime of single collective excitation. In section \ref{3}, we
discuss the principles of measurement-induced entanglement between
excitation from two remote atomic ensembles, and connection of two
pairs. The theoretical model developed in \cite{chou05} to verify
experimentally entanglement is summarized, and used to give insights
into the connection process. The experimental setup is finally
presented in section \ref{4}, together with the experimental
results. We discuss in section \ref{5} the experimental perspectives
of realizing entanglement connection.

\begin{figure*}[t]
\centering
\includegraphics[width=1.7\columnwidth]{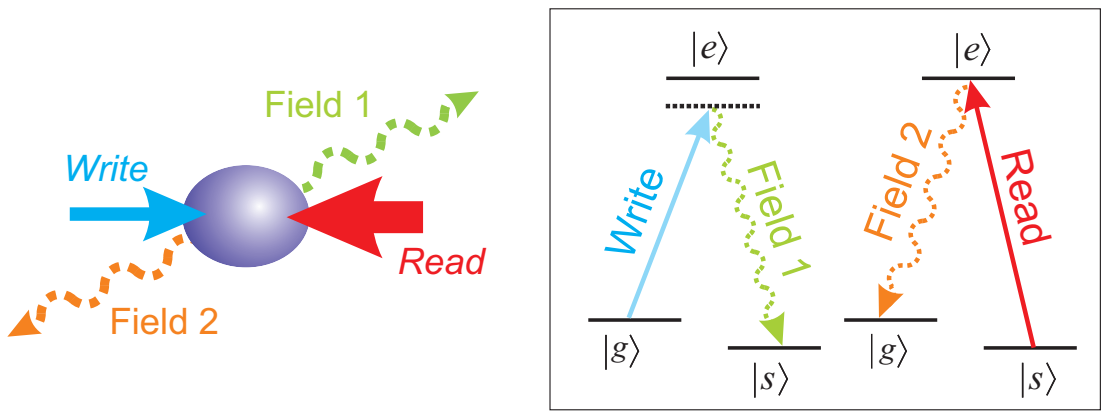}
\caption{DLCZ building block in a counter-propagating and off-axis
configuration. The inset shows the relevant atomic levels for the $%
6S_{1/2}\rightarrow 6P_{3/2}$ transition in cesium, as well as the
associated light fields. The ensemble is initially prepared in
$|g\rangle $. A weak write pulse then induces spontaneous Raman
transitions $|g\rangle \rightarrow |e\rangle \rightarrow |s\rangle
$, resulting with small probability in the emission of a photon
(field $1$, detected with at a small angle to the write beam inside
a single mode fiber) along with the storage of a collective
excitation. After a programmable delay, a strong read pulse then
maps the state of the atoms to another photonic mode, field $2$, via
$|s\rangle \rightarrow |e\rangle \rightarrow |g\rangle $.}
\label{DLCZ}
\end{figure*}

\section{Atomic ensemble in the single excitation regime}

\label{2}

In 2001, Duan, Lukin, Cirac and Zoller (DLCZ) proposed an original approach
to perform scalable long-distance quantum communications, involving atomic
ensembles, linear optics and single photon detectors \cite{duan01}. The
building block is a large ensemble of $N$ identical atoms with a $\Lambda$%
-type level configuration, as shown in figure \ref{DLCZ}. A weak
light pulse, called write pulse, with frequency close to the
$|g\rangle \rightarrow |e\rangle$ transition, illuminates the atoms
and induces spontaneous Raman scattering into a photonic mode called
field-1. For a low enough write power, such that two excitations are
unlikely to occur, the detection of a field-1 photon heralds the
storage of a single spin excitation distributed among the whole
ensemble. The joint state of the atoms and field 1 can then be
written as:
\begin{eqnarray}
|\Psi\rangle=|0_a\rangle|0_1\rangle+\sqrt{p}|1_a\rangle|1_1\rangle+O(p)
\label{psi}
\end{eqnarray}
where $|n_1\rangle$ stands for the state of the field 1 with $n$
photons and $p$ corresponds to the small probability of a single
photon scattered into field 1 by the atoms illuminated by the write
pulse. We define $|0_a\rangle\equiv\bigotimes_i^N|g\rangle_i$ and
$|1_a\rangle$ denotes a symmetric collective excitation, with
\begin{eqnarray}
|1_a\rangle=\frac{1}{\sqrt{N}}\sum^{N}_{i=1}|g\rangle_1\cdots|s\rangle_i%
\cdots|g\rangle_N.
\end{eqnarray}

A read pulse, on resonance with the $|s\rangle \rightarrow |e\rangle $
transition, can later, after a programmable delay, transfer this atomic
excitation into another photonic mode, field 2. The key element of the
protocol is that this readout can be achieved with high efficiency thanks to
a many-atom interference effect, called collective enhancement \cite%
{duanPRA,felinto05}. After readout, the ideal resultant state of the
fields 1 and 2 is:
\[
|\Phi \rangle =|0_{1}\rangle |0_{2}\rangle +\sqrt{p}|1_{1}\rangle
|1_{2}\rangle +O(p).
\]%
The photon numbers are correlated, precisely as in the case of
parametric down conversion. The lower is the excitation probability,
the better is the approximation of the non-vacuum part by a photon
pair, at the price of reduced count rates.

In our group, the optically thick atomic ensemble is obtained from
cold cesium atoms in a magneto-optical trap (MOT). At a frequency of
40 Hz, the magnetic field is switched off for 7 ms. After waiting
about 3 ms for the magnetic field to decay, sequences of writing,
reading and repumping processes are carried out for about 4 ms, with
a period of 575 ns. The weak write pulses, with a 200 $\mu$m beam
waist and linear polarization, are detuned 10 MHz below resonance.
The read pulse is orthogonally polarized to the write pulse and
mode-matched to it in a counter-propagating configuration. Both
write and read pulses are 30 ns long. Fields 1 and 2 are collected
into mode-matched fibers with a 3$^\circ$ angle relative to the
common direction
defined by write and read beams \cite{Balic05}, and with a waist of 50 $\mu$%
m defined by the backward projection of our imaging system into the sample.
Before detection, field 1 passes through a filtering stage in order to
filter out the photons that are spontaneously emitted when the atoms in the
sample go back to $|g\rangle $, without creating the desired collective
excitation.

Three parameters well characterize experimentally the system : how
well the system is in the single excitation regime, how efficient is
the retrieval of a single excitation and how long the excitation can
be stored before retrieval while preserving its coherence. The first
parameter can be determined by a measurement of the suppression of
the two-photon component of the field 2 obtained from the retrieval
of the excitation. Suppression below 1$\%$ of the value for a
coherent state has been reported in our system \cite{laurat06}. The
ability to efficiently retrieve the excitation is also critical. The
probability to have a photon in field 2 in a single spatial mode at
the output of the atomic ensemble once an event has been recorded
for field 1 can be as high as 50$\%$, leading to a probability
around 12.5 $\%$ for having a detection event \cite{laurat06}. Last
but not least, the writing and retrieval processes can be separated
by a programmable delay. As this delay is increased, the above two
quantities decay in a typical time scale around 10 to 20 $\mu$s. The
principal cause for this finite coherence time is the residual
magnetic field that inhomogeneously broadens the ground state levels
of the atomic samples. Detailed theoretical and experimental studies
of the decoherence have been reported in Refs.
\cite{felinto05,deRiedmatten06,felinto06}.

\section{Measurement-induced entanglement and connection of atomic ensembles}

\label{3} Starting from this building block, DLCZ proposed in their
seminal paper to generate and store entanglement for excitation in
two remote ensembles and then to connect two pairs. This section
presents these measurement-induced schemes, which rely on quantum
interference in the detection of a photon emitted by one of the
ensembles. After establishing entanglement, directly or via
connection, a difficult experimental task is
to prove the entanglement \cite{vanEnk06}. A robust model developed in \cite%
{chou05} is then presented.

\begin{figure*}[t]
\centering
\includegraphics[width=1.5\columnwidth]{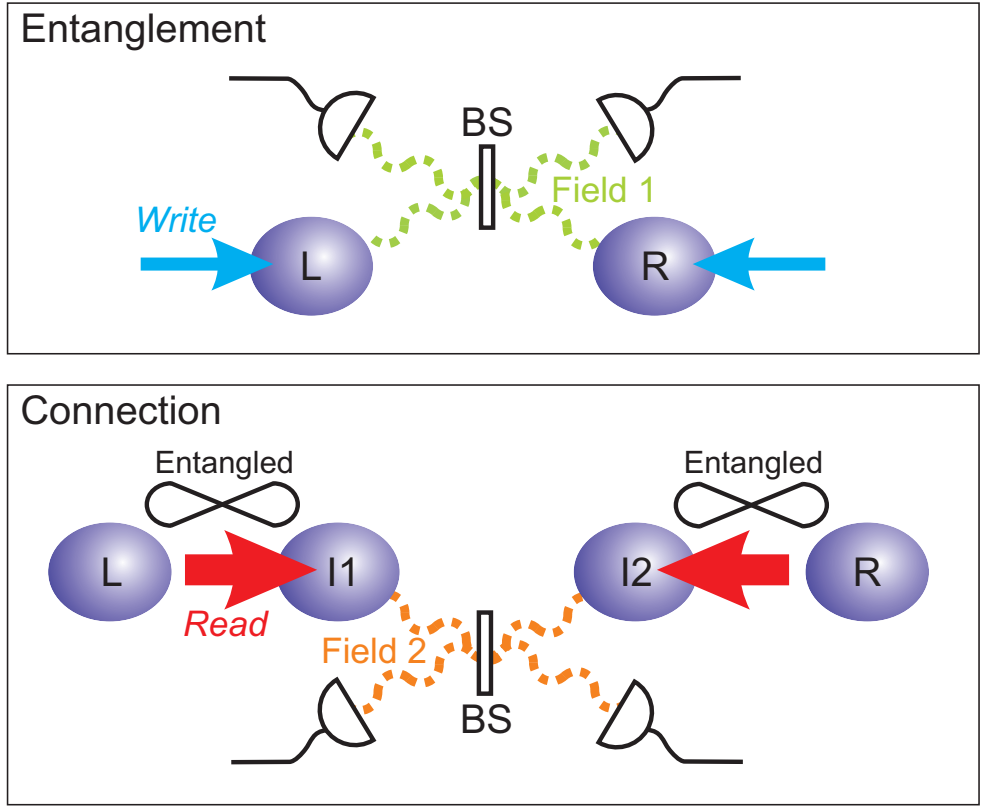}
\caption{Setup for (a) entanglement generation between two atomic
ensembles and (b) entanglement connection between two pairs. In both
cases, the entanglement results from the interference of photonic
modes at a 50/50 beamsplitter, with outputs directed towards
single-photon detectors. The photonic modes consist in (a) field 1
modes from both ensembles or (b) field 2 modes after reading of one
ensemble of each pair.} \label{principles}
\end{figure*}
\subsection{Entanglement between two ensembles}

Let us consider now two atomic ensembles, for which fields 1 are
superposed on a 50/50 beamsplitter, in an indistinguishable way,
with the outputs directed towards two single-photon detectors
(Figure \ref{principles}). The detection of a field 1 photon from
either of the two ensembles results in an entangled state with one
excitation shared coherently between the two ensembles. Such
measurement-induced entanglement has been demonstrated recently in
our group for excitation stored in two atomic ensembles separated by
3 meters \cite{chou05}.

In more details, after two write pulses are sent into the two ensembles
simultaneously, the scattered fields 1 and ensembles are in the product
state :
\begin{eqnarray}
|\Psi _{LR}\rangle  &\propto &[|0_{a}\rangle |0_{1}\rangle +e^{i\beta _{L}}%
\sqrt{p}|1_{a}\rangle |1_{1}\rangle +O(p)]_{L} \nonumber\\
&&\quad  \otimes \lbrack |0_{a}\rangle |0_{1}\rangle +e^{i\beta
_{R}}\sqrt{p}|1_{a}\rangle |1_{1}\rangle +O(p)]_{R}.
\end{eqnarray}%
Here $\beta _{R}$ and $\beta _{L}$ correspond to overall propagation
phases determined by the write pulses. Detection of a photon in
either detector then projects the state of the ensembles as follows,
in the ideal case where higher-order terms are neglected:
\begin{eqnarray}
\rho _{LR}^{^{\prime }} &=&Tr_{1_{L}1_{R}}[\rho (\frac{1}{\sqrt{2}}%
(a_{1_{L}}\pm e^{i\theta }a_{1_{R}})|\Psi _{LR}\rangle
)]\nonumber\\&=&|\Psi _{LR}^{^{\prime }}\rangle\langle\Psi
_{LR}^{^{\prime
}}|  \nonumber  \label{pm} \\
\quad\textrm{with}&&|\Psi _{LR}^{^{\prime
}}\rangle=\frac{1}{\sqrt{2}}(|0_{a}\rangle _{L}|1_{a}\rangle _{R}\pm
e^{i\eta }|1_{a}\rangle _{L}|0_{a}\rangle _{R}).
\end{eqnarray}%
$\rho (|\Psi \rangle )\equiv|\Psi \rangle \langle \Psi |$,
$Tr_{1_{L}1_{R}} $ stands for tracing over the states of fields
$1_{L}$ and $1_{R}$, $a_{1_{L}}$ and $a_{1_{R}}$ are the
annihilation operators associated with fields $1_{L}$ and $1_{R}$,
$\theta=\theta_R-\theta_L $ the difference phase shift between the
two field 1 paths from the ensemble to the beamsplitter, and finally
the overall phase $\eta =(\beta _{L}-\beta _{R})+(\theta _{L}-\theta
_{R}) $. This phase $\eta $ is the sum of the phase difference of
the write beams at the L and R ensembles and the phase difference
acquired by fields 1 in propagation from the ensembles to the
beamsplitter. To achieve entanglement, this phase has to be kept
constant. In order to meet this stringent and challenging
requirement in the initial demonstration reported in \cite{chou05},
the different phases have been independently controlled and actively
stabilized by using auxiliary fields. Finally, the $\pm $ sign in
Eq. \ref{pm} comes from the $\pi $ phase difference between the two
outputs of a beam splitter: depending on which detector records the
heralding event, two different entangled states are generated, and
stored for subsequent utilization.

\subsection{Entanglement connection}

When two pairs of atomic ensembles are prepared in such an entangled
state (figure \ref{principles}), one can connect the pairs by
sending strong read pulses into one ensemble of each pair. The
fields 2 resulting from this readout are then brought to
interference at a 50/50 beamsplitter. Again, a single click on
either detector prepares the remaining ensembles in an entangled
state \cite{duan01}.

After independent preparation of entanglement for the pairs $\{L,I1\}$ and $\{%
R,I2\}$ and perfect reading of the states of the ensembles I1 and
I2, the joint state of the fields 2 and the ensembles can be
written, neglecting higher order terms:
\begin{eqnarray}
|\Psi _{L,R,2_{I1},2_{I2}}\rangle &=&\frac{1}{2} [|0\rangle
_{2_{I2}}|1_{a}\rangle _{R}\pm e^{i\zeta _{R,I2}}|1\rangle
_{2_{I2}}|0_{a}\rangle _{R})] \nonumber\\
&& \otimes \lbrack |0\rangle _{2_{I1}}|1_{a}\rangle _{L}\pm
e^{i\zeta _{L,I1}}|1\rangle _{2_{I1}}|0_{a}\rangle _{L})]
\end{eqnarray}%
where the phases resulting from the entanglement generation and the
readout process are given by $\zeta _{i,Ij}=(\beta _{Ij}-\beta
_{i})+(\theta_{Ij}-\theta _{i})+\delta_{Ij}$, with $\delta_{Ij}$ the
phase of the read beam at the $Ij$ ensemble. Fields $%
2_{I1}$ and $2_{I2}$ are then mixed on a 50/50 beamsplitter, and
detection of a photon in either detector projects the remaining two
ensembles $L$ and $R$ into:
\begin{eqnarray}
\rho _{LR} &=&Tr_{2_{I1}2_{I2}}[\rho
(\frac{1}{\sqrt{2}}(a_{2_{I1}}\pm e^{i\gamma }a_{2_{I2}})|\Psi
_{L,R,2_{I1},2_{I2}}\rangle )]\nonumber\\
\end{eqnarray}%
which can be written as
\begin{eqnarray}\label{final}
\rho _{LR}&=&\frac{1}{2}|0\rangle \langle 0|+\frac{1}{2}|\Phi
_{L,R}\rangle \langle \Phi _{L,R}|\nonumber\\\quad
\textrm{with}&&\quad |\Phi _{L,R}\rangle =|0\rangle _{L}|1\rangle
_{R}\pm e^{i\xi }|1\rangle _{L}|0\rangle _{R}
\end{eqnarray}%
where $\xi =\zeta _{R,I2}-\zeta _{L,I1}+\gamma $. This overall phase
is the sum of the phase difference for entanglement generation for
each pair, the phase difference between the two read beams up to the
two ensembles and the phase difference of the generated fields 2
from the ensembles to the beamsplitter.

The vacuum part comes from the probability of reading the two
excitations at the same time, leaving no remaining excitation in the
system. In the ideal case, the connection succeeds 50$\%$ of the
time. Let us underline also that, significantly, the absolute phases
do not necessarily need to be stabilized to succeed in the
connection. Only the overall phase $\xi$ must be kept constant. This
feature is exploited in the proposed experimental setup, where
passive stability is found to be enough to meet this requirement.

The generated state given by Eq. \ref{final} is what DLCZ called an
\textquotedblleft effective maximally-entangled
state\textquotedblright (EME) as any state of this form would be
purified to a maximally entangled state in the proposed
entanglement-based communication scheme \cite{duan01}. The vacuum
coefficient only influences the success probability, but not the
overall fidelity of the long-distance communication. This important
feature is known as \textquotedblleft built-in
purification\textquotedblright.

\begin{figure*}[t]
\centering
\includegraphics[width=1.8\columnwidth]{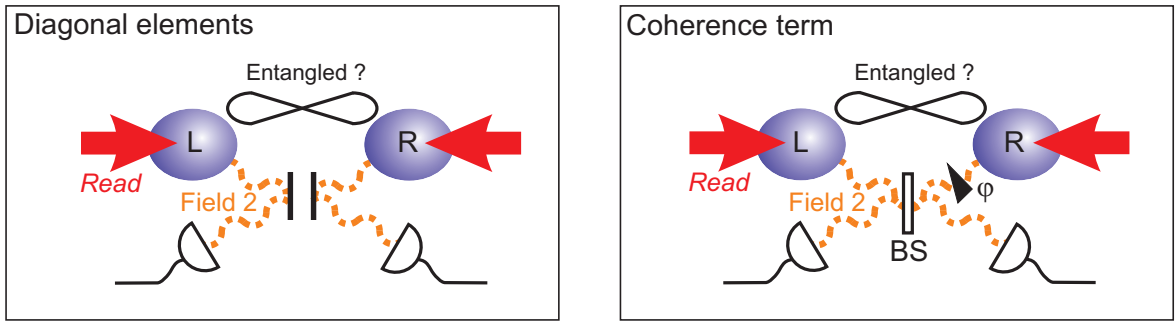}
\caption{Procedure for verifying entanglement between two atomic
ensembles. The atomic state is mapped to photonic modes via
simultaneous strong read pulses and quantum tomography of the
generated fields 2 is performed. For this purpose, fields 2 are
detected independently (diagonal elements) or in a 50/50
beamsplitter configuration where the phase of one of the paths is
scanned (coherence term).} \label{measurement}
\end{figure*}

\subsection{Experimental verification of entanglement}

\label{3_measure} To experimentally verify the entanglement between
the two atomic ensembles, L and R as sketched in figure
\ref{measurement}, a solution is to map the delocalized atomic
excitation into a field state by applying simultaneous strong read
pulses. For perfect state transfer, the entangled state of the atoms
would be mapped to an ideal entangled state of the two photonic
modes.

However, the presence of various noises, the vacuum contribution
(coming from a finite retrieval efficiency or also a finite success
probability in the case of the swapping), as well as higher order
terms, have to be taken into account. In order to prove
experimentally the generation of
entanglement at the atomic level, our group has developed in Ref. \cite%
{chou05} a robust, model-independent determination of entanglement
based upon quantum tomography of the fields 2. As entanglement
cannot be increased by local operations on either of the two
ensembles, the entanglement for the state of the ensembles will
always be greater than or equal to that measured for the light
fields. The model consists of reconstructing a reduced density
matrix, $\rho $, obtained from the full density matrix by
restricting it to the subspace where no more than one photon
populates each mode. It can be shown that this reduced density
matrix exhibits less or equal entanglement than the full one. The
model will thus lead to a lower bound of the entanglement, enabling
an unambiguous determination of the presence of entanglement, at the
price of eventually underestimating its actual magnitude.

The reduced density matrix can be written as:
\begin{eqnarray}
\rho=\frac{1}{P}\left(
\begin{array}{cccc}
p_{00} & 0 & 0 & 0 \\
0 & p_{01} & d & 0 \\
0 & d^{*} & p_{10} & 0 \\
0 & 0 & 0 & p_{11}%
\end{array}
\right)
\end{eqnarray}
in the photon-number basis $|n\rangle |m\rangle$, with $\{n,m\}=\{0,1\}$%
. $p_{ij}$ is the probability to find i photons in mode $2_L$ and j in mode $%
2_R$, $d$ is the coherence term between the $|1\rangle|0\rangle$ and $%
|0\rangle|1\rangle$ states, and $P=p_{00}+p_{01}+p_{10}+p_{11}$.
From this density matrix, one can calculate the concurrence $C$,
which is a monotone measurement of entanglement \cite{wooters98}:
\begin{eqnarray}
P.C=max(2|d|-2\sqrt{p_{00}p_{11}},0)
\end{eqnarray}
Let us underline, as $d^2\leq p_{10}p_{01}$, a necessary requirement for $C>0
$ is that there is a suppression of two-photon events relative to the square
of the probability of single photon events for the fields 2 : $h\equiv
p_{11}/(p_{10}p_{01})<1$.

Experimentally, the density matrix can be reconstructed by using two
different configurations, as sketched in figure \ref{measurement}.
The diagonal elements are determined by measuring individual
statistics, i.e. by detecting independently each field. The
coherence term can be measured by combining the fields 2 on a 50/50
beamsplitter and recording the count rate as a function of the phase
difference between them. This results in an interference fringe with
a visibility $V$. It can been shown that $d\simeq
V(p_{10}+p_{01})/2$. Together, this two-stage measurement gives
access to the concurrence $C$.

\subsection{Entanglement connection revisited}

The principle of entanglement connection has been explained
previously in the ideal case where higher order terms and vacuum
contributions are neglected. Let us consider now the more general
case, which can be described by the previous approach. We consider
two pairs of entangled ensembles and consider that the fields 2
after reading can be described by the same density matrix
$\rho^{\prime }$ with diagonal elements $p_{ij}^{\prime }$. The
relevant question is now what will be the expression of $\rho$, the
reduced density matrix for the fields 2 of the two remaining
ensembles after the connection.

Let us assume that $p_{10}^{\prime }=p_{01}^{\prime }$. To later
normalize the events conditioned on swapping, one needs to first
determine the probability to have a click heralding the connection
at one output of the beamsplitter. To the first order, this quantity
can be written as:
\begin{eqnarray}
p^{\prime }=2\times\frac{1}{2}p_{10}^{\prime }=p_{10}^{\prime }.
\end{eqnarray} The factor $1/2$ corresponds to the $50\%$ chance
that the photon be reflected or transmitted at the beamsplitter,
while the factor 2 results from the symmetry of the scheme where the
photon can come from either ensemble.

One can then evaluate, after the reading of the two remaining
ensembles, the probability to have one photon for one mode and zero
for the other, when a swap event has been detected:
\begin{eqnarray}
p_{10} &=&p_{01}=\frac{1}{2}%
(p_{10}^{\prime\,2 }+p_{11}^{\prime }p_{00}^{\prime }+p_{11}^{\prime
}p_{10}^{\prime })/p^{\prime }\sim \frac{1}{2}p_{10}^{\prime }.
\end{eqnarray}
The terms inside the parenthesis correspond to one photon in mode
$2_L$ and zero in $2_R$ (or the other way round), and all the other
combinations for $2_{I1}$ and $2_{I2}$  which can give a swapping
event. The final factor $1/2$ comes from the fact, already
established before in the ideal case, that the swapping succeeds, to
the first order, 50$\%$ of the time.

Finally, in a similar way, the probability to have one photon in
each mode is given by:
\begin{eqnarray}
p_{11} &=&\frac{1}{2}p_{11}^{\prime }(p_{11}^{\prime
}+2p_{10}^{\prime })/p^{\prime }\sim p_{11}^{\prime }.
\end{eqnarray}
The main feature which appears here is that the weight of the two
photon component stays the same, while the single-photon component
is divided by two. As a result, if one calculates for the connected
pairs the new suppression $h$ of two-photon events relative to the
square of the probability for single photon events as a function of
the initial $h^{\prime }$ for each entangled pair: $h\sim4h^{\prime
}$. This result points out the difficulty which could arise in the
experimental demonstration of entanglement connection: one needs to
start with atomic ensembles entangled with a very low two-photon
component, at the price of low count rates and statistics.

\section{Experimental setup and measurement results}

\label{4} In this section, we present a scheme that permits us to
investigate entanglement connection between two pairs of atomic ensembles,
without the requirement of any active phase stabilization. Experimental
results are finally given.
\begin{figure*}[t]
\centering
\includegraphics[width=1.6\columnwidth]{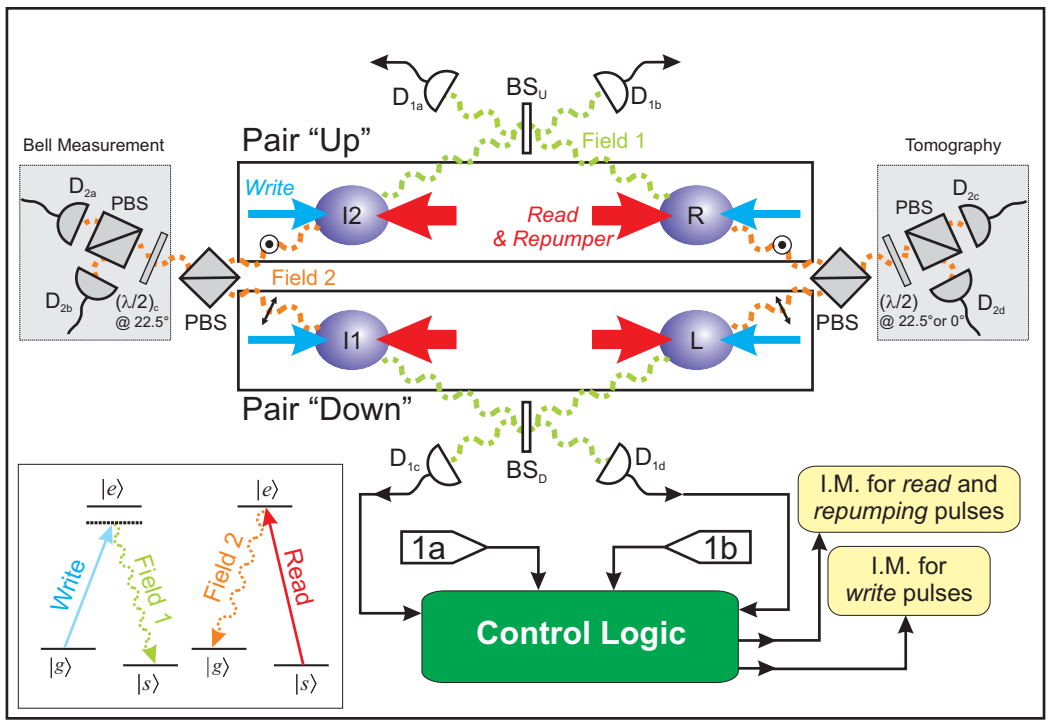}
\caption{Experimental setup. For each pair, Up and Down, the ensembles are
separated by 3 meters. Fields $1_{I2}$ and $1_{R}$ from pair Up are brought
to interference at a 50-50 beam splitter ($BS_{U}$). A photo-detection event
at either detector $D_{1a}$ or $D_{1b}$ heralds entanglement between the
collective excitation in $I2$ and $R$. The Down pair is prepared in a
similar fashion via events at $D_{1c},D_{1d}$. A heralding detection event
triggers the control logic to gate off the light pulses going to the
corresponding ensemble pair by controlling intensity modulators (\textit{IM}%
). The atomic state is thus stored while waiting for the second ensemble
pair to be prepared. After both pairs of ensembles have been prepared, the
control logic releases strong read pulses. Fields $2_{I2}$ and $2_{I1}$ ($%
2_{R}$ and $2_{L}$) are combined with orthogonal polarizations on
polarizing beam splitters. Fields $2_{I2}$ and $2_{I1}$ are detected
with the half-wave plate $(\protect\lambda/2)_{c}$ at 22.5$^\circ$,
which is equivalent to a 50/50 beamsplitter configuration. The
fields 2 from the remaining ensembles are characterized
conditionally on a detection event heralding the connection. The two
configurations of figure \protect\ref{measurement} correspond to two
different angles, 0$^\circ$ and 22.5$^\circ$, of the half-wave plate
$(\protect\lambda/2)$.} \label{setup}
\end{figure*}

\subsection{Experimental setup}

The experimental setup is depicted in Figure \ref{setup}. Two
parallel pairs of atomic ensembles are first prepared independently,
following the measurement-induced method detailed in section
\ref{3}. This preparation stage is sped up by real-time conditional
control \cite{felinto06,chou07}: a detection event at either pair
triggers intensity modulators that gate off all lasers pulses going
to the corresponding pair of ensembles, thereby storing the
associated state. After successfully preparing both pairs,
strong read pulses are sent into the ensembles. The fields $2_{I1}$ and $%
2_{I2}$ are brought to interfere and a detection event on either
detector heralds the connection process. Thanks to the conditional
control, a 20-fold enhancement is obtained in the probability to
establish the connection, leading to a rate of connection around
4Hz. Depending on the combinations of field 1 and field 2 detector
clicks, two different entangled states are generated for the two
remaining ensembles, denoted by $+$ and $-$, with a $\pi$
phase-shift between them.

As pointed out before, the process of connection between the two
remaining ensembles, which never interacted in the past, only
requires the stability of the relative phase $\xi$ over trials. This
overall phase is defined as the phase difference between the
absolute phase of all the paths (write beams, field 1, read pulses,
and field 2 on the connection side) for the upper pair and the ones
for the lower pair. Instead of actively stabilizing all individual
phases as it was performed in \cite{chou05} where two ensembles were
involved, this requirement is fulfilled in our setup by exploiting
the passive stability between two independent polarizations
propagating in a single interferometer \cite{chou07}. All the paths
for the upper and lower pairs are common, except inside a small
interferometer where orthogonal polarizations are separated to
define the two ensembles on each side \cite{chouthesis}. Operation
over more than 24 hours is possible without any adjustment as the
phase does not change by more than a few degrees. As a result, no
active phase stabilization is required, simplifying significantly
the experimental investigation of the connection process. Note that
although the present configuration is sufficient to demonstrate the
principle of the connection, an experiment where the final pair of
ensembles L and R are distant, as in Fig 2b, would require active
stabilization of the various phases [9], since in that case all the
paths would be distinct. Our configuration for passive stability is
better suited to the case of parallel chains of ensembles, as in the
original proposal of DLCZ.

\begin{table}[t]
\caption{Diagonal elements of the density matrix $\protect\rho$
deduced from the records of photo-electric counts, for the two
different states after connection, denoted $+$ and $-\,$. These
values are obtained by considering unit detection efficiency. Errors
bars correspond to statistical errors.} \label{pij}{\footnotesize
\begin{ruledtabular}
\begin{tabular}{lcc}
Probability&+&-\\
$p_{00}$&$0.949\pm0.003$&$0.948\pm0.003$\\
$p_{10}$&$( 1.97\pm0.05)\times 10^{-2}$&$(1.99\pm0.05)\times 10^{-2}$\\
$p_{01}$&$(3.06\pm0.06)\times 10^{-2}$&$(3.16\pm0.06)\times 10^{-2}$\\
$p_{11}$&$(4.1\pm0.7)\times 10^{-4}$&$(4.9\pm0.8)\times 10^{-4}$\\
\end{tabular}
\end{ruledtabular}
}
\end{table}

\subsection{Characterization of the states generated upon connection}

The generated state is analyzed by using the tomography technique explained
in section \ref{3_measure}. Conditioned upon a connection event, the density
matrix $\rho$ of the fields 2 is reconstructed following the two required
steps: the measurement of the diagonal elements and the determination of the
coherence term.

Table \ref{pij} gives the measured diagonal elements deduced from the
records of photo-electric counts, for both generated states, after a
connection event. Unit detection efficiency is assumed, which can only lead
to a smaller value for the concurrence than the actual field concurrence for
finite detection efficiency. From theses values, one can deduce the suppression $%
h$ of the two-photon events relative to the square of the
probability for single-photon events. We find $h_+=0.7\pm0.1<1$ and
$h_-=0.8\pm0.1<1$. From independent measurements, we inferred the
$h^{\prime } $ parameter for each pair before connection to be
$h^{\prime }=0.20\pm0.05$. The experimentally determined values of
$h$ are thus consistent with the expression $h=4h^{\prime }$
established previously. As pointed out before, this relation arises
from the intrinsic feature that the connection
succeeds only 50$\%$ of the time. This can be seen in the quantities $%
2p_{01}\sim4\%$ and $2p_{10}\sim6\%$, which should be equal to half
the retrieval efficiency. The retrieval efficiencies (including
detection) independently measured for each ensemble were both around
10$\%$.

\begin{figure*}[htpb!]
\centering
\includegraphics[width=1.85\columnwidth]{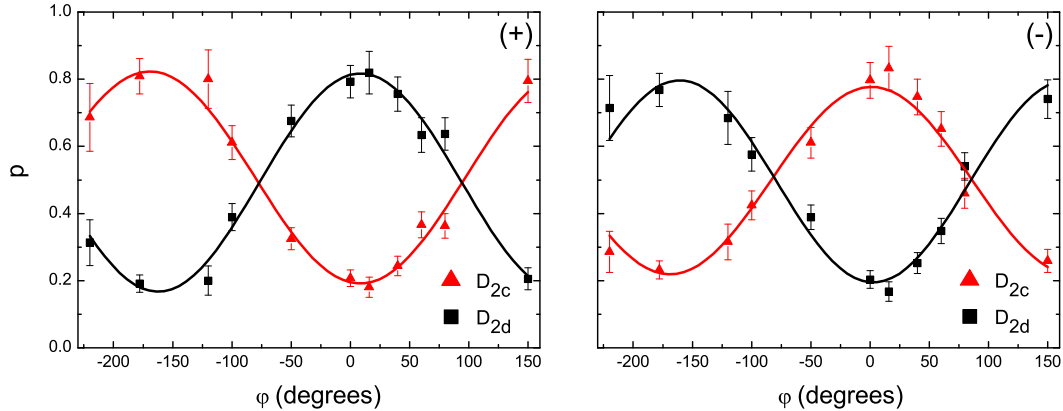} \vspace{-5mm}
\caption{Coherence between the two atomic ensembles $L$ and $R$
induced by the connection event. $p$ is the probability after
connection to have a detection event on either detectors $D2c$ and
$D2d$ when the fields $2_L$ and $2_R$ interfere, as a function of
the phase $\protect\varphi$. For each phase setting, data are
acquired for 30 minutes, each atomic state being generated overall
at about 2Hz. Errors bars correspond to statistical errors. }
\label{fringe}
\end{figure*}
\begin{figure}[t]
\centering
\includegraphics[width=0.9\columnwidth]{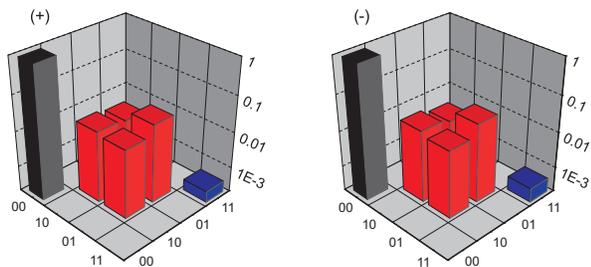}
\caption{Reconstructed density matrix for both generated states, at
the detector location.} \label{matrix}
\end{figure}

In order to access the coherence term, figure \ref{fringe} shows the
probability to have a detection event on either output of the beam
splitter, normalized to the sum of these events, as a function of
the phase-shift between the fields $2_{L}$ and $2_{R}$. Practically,
the relative phase is scanned by adjusting the phases of the two
classical read beams via birefringent waveplates. The visibilities
are found to be $V_{+}=64\pm 3\%$ and $V_{-}=59\pm 3\%$. A simple
model \cite{chou07} predicts for our excitation probability a
visibility equal to $65\%\pm 10\%$. By taking into account the
measured overlap of the photon wavepacket for fields 2 deduced from
a two-photon interference \cite{chou07}, $ 0.90\pm0.05$, the
expected visibility can be roughly estimated to be $55\pm 10\%$ if
all the reduction is attributed to a non-perfect overlap. In the
absence of conditioning, the visibility drops to near zero, the
residual visibility  (below $3\%$) being explained by finite
polarization extinction ratio in our setup. This result demonstrates
for the first time the creation of coherence between two atomic
ensembles which never interacted in the past. The reconstructed
density matrices are shown in figure \ref{matrix}.

With these data in hand, the concurrences $C$ can be estimated for both
states:
\begin{eqnarray}
C_+&=&max\left(-(7\pm4)\times 10^{-3},0\right)=0 \\
C_-&=&max\left(-(1.3\pm0.4)\times 10^{-2},0\right)=0.
\end{eqnarray}
These values show finally the absence of entanglement, or at least,
that our entanglement measurement, which provides a lower bound of
the atomic entanglement, cannot detect entanglement in this
particular case. One can correct from detection efficiencies and
propagation losses \cite{chou05}, but any zero concurrence will stay
zero by this correction. The $h$ values confirm anyway that the
connected systems are barely in the regime where the two-photon
events are suppressed relative to single photon events. One needs to
start with smaller $h^{\prime }$ for the initial pairs. $h^{\prime
}$ as low as 0.05 can be obtained routinely for each pair in our lab
but the count rate to characterize the connection would be
prohibitively low.

\section{Discussion and perspectives}

\label{5}

In summary, we have presented a possible scheme to demonstrate entanglement
connection between atomic ensembles which never interacted in the past. Such
striking capability is a critical requirement for the future development of
elaborate quantum networks. Our investigation has shown for the first time
the creation of coherence upon the connection process. This result validates
our proposed setup, in particular its passive phase stability, and
constitutes a significant step towards the entanglement connection of matter
systems.

To finally generate and prove entanglement connection between the
remaining ensembles, very stringent condition on the suppression of
the two-photon component needs to be satisfied, at the sacrifice of
the count rate in our current setup. Overall, the figure of merit of
any elaborate experiment is the product of the probability to
prepare the entangled state at each write pulse and the coherence
time. Improvements in these two directions have to be explored. The
first one can be addressed by, for instance, multiplexing the atomic
ensembles. One can imagine to use spatially-resolving detectors,
namely array of single-photon detectors, and adaptive optical
systems to reconfigure the optical interconnects. Improving the
coherence time is a second critical direction as more elaborate
protocols are involved. It would require better nulling of the
residuals
magnetic fields and also the use of improved trapping techniques \cite%
{felinto05} like a large dipole-trap, as a magneto-optical trap will
be rapidly limited by the diffusion of the atoms outside the
excitation region. An increase by two orders of magnitude, from tens
of $\mu s$ to $ms$, would enable for instance to demonstrate the
entanglement connection in our current setup in a few hours of data
taking. All together, these improvements would enable deeper
investigation of experimental quantum networking, and will
definitely lead to fruitful insights into the distribution and
processing of quantum information.

\section*{Acknowledgements}

We gratefully acknowledge our ongoing collaboration with S.J. van Enk. This
research is supported by the Disruptive Technologies Office (DTO) and by the
National Science Foundation (NSF). J.L. acknowledges financial support from
the European Union (Marie Curie fellowship). H.D. acknowledges support as
Fellow of the Center for the Physics of Information at Caltech.

\end{document}